\documentclass[12pt]{article}
\usepackage{amssymb}
\usepackage[dvips]{epsfig}

\newcommand{\be}{\begin{equation}}
\newcommand{\ee}{\end{equation}}
\def\bea{\begin{eqnarray}}
\def\eea{\end{eqnarray}}

\newcommand{\stuck}{St$\ddot{u}$ckelberg }
\newcommand{\bn}{\begin{eqnarray}}
\newcommand{\en}{\end{eqnarray}}

\newcommand{\p}{\partial}

\newcommand{\nn}{\nonumber}

\newcommand{\no}{\noindent}

\newcommand{\tk}{\tilde{K}}

\newcommand{\llh}{\p^{\mu}\p^{\nu} h_{\mu\nu}}
\newcommand{\dk}{\p^{\mu}\p^{\nu} K_{\mu\nu}}

\newcommand{\s}{\,\,\,\,}
\def\bea{\begin{eqnarray}}
\def\eea{\end{eqnarray}}

\newcommand{\beq}{\begin{eqnarray}}
\newcommand{\eeq}{\end{eqnarray}}
\begin{document}

\title{\textbf{Massive spin-2 particles via embedment of the Fierz-Pauli equations of motion}}
\author{D. Dalmazi\footnote{dalmazi@feg.unesp.br},
A.L.R. dos Santos\footnote{alessandroribeiros@yahoo.com.br}, E.L. Mendon\c ca\footnote{elias.fis@gmail.com} \\
\textit{{UNESP - Campus de Guaratinguet\'a - DFQ} }\\
\textit{{Avenida Dr. Ariberto Pereira da Cunha, 333} }\\
\textit{{CEP 12516-410 - Guaratinguet\'a - SP - Brazil.} }\\}
\date{\today}
\maketitle

\begin{abstract}

Here we obtain alternative descriptions of massive spin-2
particles by an embedding procedure of the Fierz-Pauli equations
of motion. All models are free of ghosts at quadratic level
although most of them are of higher order in derivatives. The
models that we obtain can be nonlinearly completed in terms of a
dynamic and a fixed metric. They include some $f(R)$ massive
gravities recently considered in the literature. In some cases
there is an infrared (no derivative) modification of the
Fierz-Pauli mass term altogether with higher order terms in
derivatives. The analytic structure of the propagator of the
corresponding free theories is not affected by the extra terms in
the action as compared to the usual second order Fierz-Pauli
theory.

\end{abstract}

\newpage

\section{ Introduction}

Massive gravity has become an area of intense work in last 5
years, see the review works \cite{hinter,rham} and references
therein. Part of the motivation has an earlier origin, they come
from the experimental data of the supernova team
\cite{super1,super2}, which indicate an accelerated expansion of
the universe at large distances. A tiny mass for the graviton
would certainly diminish the gravitational interaction at large
distances which might avoid the introduction of dark energy. From
the theoretical standpoint the motivation comes from the new
formulations of massive gravity of \cite{rg,rgt}, see also
\cite{hr1,hr2}, which overcome two important obstacles. Namely,
the Boulware-Deser ghost \cite{db} and the van
Dam-Veltman-Zakharov mass discontinuity \cite{vdv,zak} which is
solved along the lines of \cite{vain}.

The theories \cite{rgt,hr1} are built up on the top  of the second order (in derivatives) paradigmatic
Fierz-Pauli (FP) theory \cite{fp}. Even more recent works \cite{hinter2,ky,rmt13}  searching for alternative
kinetic terms for massive gravity, reduce to the FP model at linearized level. On the other hand, in order to
accommodate a larger class of stable cosmological solutions, higher order  modifications of \cite{hr1} and
\cite{rgt} have been recently considered \cite{no,nos,cds,kno,cs}. So it is natural to ask for possible higher
order alternatives to the FP paradigm.

Here we start with the second order FP theory and add quadratic
terms in its equations of motion and its derivatives as a
technique to generate higher order dual theories which have the
same particle content of the massive FP model at quadratic level.
Namely, a massive spin-2 particle (in $D=4$) and no ghosts. By
requiring that the FP equations of motion are not only embedded
but in fact follow from the higher order model, we end up with the
condition (\ref{det}) which has several solutions. In the next
section we write down explicitly four families of solutions. It
turns out that we can always find a nonlinear completion of those
quadratic (free) theories in terms of a dynamic $g_{\mu\nu}$ and a
fixed $g_{\mu\nu}^{(0)}$ metric. Those theories might be
considered as alternative starting points for the addition of
possible extra terms required for the absence of ghosts at
nonlinear level. In particular, a subset of the models
(\ref{snl1}) have been already reformulated as ghost free theories
in \cite{cds,cs} by adding nonlinear non derivative terms as in
\cite{rg,rgt}. In section 3 we reinterpret our results in terms of
the analytic structure of the propagator while in section 4 we
explain the essential differences of our embedment and the Noether
gauge embedment where the analytic structure of the propagator is
modified. In section 5 we draw our conclusions.

\section{Embedding the Euler tensor}

Although we use in this work mostly a symmetric tensor $h_{\mu\nu}=h_{\nu\mu} $ , it is convenient for the
introduction of the embedding idea to recall some results of the works \cite{rank2,spec} where a generic
nonsymmetric tensor $e_{\mu\nu} = h_{\mu\nu} + B_{\mu\nu}$ , with $B_{\mu\nu} = -B_{\nu\mu}$, is employed. The
nonsymmetric FP model (NSFP) is given by\footnote{Throughout this work we use $\eta_{\mu\nu} = (-,+,\cdots,+)$.}

\newpage

\bea {\cal L}_{NSFP} &=& {\cal L}_{FP}[h_{\mu\nu}] + \frac{m^2}2 B_{\mu\nu}^2  \nn\\
&=&  \frac 12 h^{\mu\nu} \left( \Box - m^2 \right) h_{\mu\nu} - \frac 12 h \left( \Box - m^2 \right)h + (
\p^{\mu}h_{\mu\nu} )^2 - \p^{\mu} \, h \, \p^{\nu}\, h_{\mu\nu} + \frac{m^2}2 B_{\mu\nu}^2 \nn\\ \label{nsfp}
\eea

We have concluded in \cite{rank2} that besides the NSFP theory, there are two extra one-parameter families of
models in $D=4$ describing massive spin-2 particles via a nonsymmetric tensor, namely\footnote{The model ${\cal
L}(a_1)$ at $a_1=-1/4$ has appeared before in \cite{ms}},

\be {\cal L}(a_1) = {\cal L}_{NSFP} + \left( a_1 - \frac 14
\right) \left( \p^{\mu}h_{\mu\nu} + \p^{\mu}B_{\mu\nu} - \p_{\nu}
\, h \right)^2 = {\cal L}_{NSFP} + \left( a_1 - \frac 14 \right)
\left(\frac{ \p^{\mu}\tk_{\mu\nu}}{m^2} \right)^2 \quad ,
\label{la1k} \ee

and

\bea {\cal L}(c) &=& {\cal L}_{NSFP} - \frac 13 \left(
\p^{\mu}h_{\mu\nu} + \p^{\mu}B_{\mu\nu} - \p_{\nu} \, h \right)^2
- m^2 \frac{(1+c)}2 \, h^2 \nn\\ &=& {\cal L}_{NSFP} -
\frac{1}{3\, m^4} \left( \p^{\mu}\tk_{\mu\nu} \right)^2 -
\frac{(1+c)}{18 m^2 } \left\lbrack \tk + \frac{2}{m^2}
\p^{\mu}\p^{\nu}\tk_{\mu\nu} \right\rbrack ^2 \nn\\ \label{lnfp}
\eea

\no where $a_1$ and $c$ are arbitrary real constants and $\tk=\eta_{\mu\nu}\tk^{\mu\nu}$ is the trace of the
Euler-tensor $\tk^{\mu\nu}$ of the NSFP theory which can be written in terms of the Euler tensor $K^{\mu\nu}$ of
the usual (symmetric) Fierz-Pauli theory :

\be \tk^{\mu\nu} \equiv  \frac{\delta \, S_{NSFP}}{\delta \,
e_{\mu\nu}} = K^{\mu\nu} + m^2 \, B^{\mu\nu} \quad , \label{tk}
\ee

\no with

\bea K^{\mu\nu} &\equiv &  \frac{\delta \, S_{FP}}{\delta \,
e_{\mu\nu}} \nn\\
&=& \left(\Box - m^2 \right) h^{\mu\nu} + \p^{\mu}\p^{\nu}\, h -
\p^{\mu}\p_{\alpha}h^{\alpha \nu} - \p^{\nu}\p_{\alpha}h^{\alpha
\mu} + \eta^{\mu\nu} \left( \p_{\alpha}\p_{\beta}h^{\alpha\beta} -
\Box \, h + m^2 \, h \right)  \label{K} \eea

Although the antisymmetric field $B_{\mu\nu} $ is completely decoupled in  (\ref{nsfp}), its presence allows us
to figure out that the families ${\cal L}(a_1)$ and ${\cal L}(c)$ differ from the NSFP theory by the addition of
quadratic terms in its equations of motion, see (\ref{la1k}) and (\ref{lnfp}). Such feature automatically
guarantees that the NSFP equations of motion $\tk^{\mu\nu} = 0$ minimize also the actions $S(a_1)$ and $S(c)$.
However, in order to make sure that the addition of the square terms does not change the particle content of the
original theory, we must go beyond the embedding and check that the new equations of motion are completely
equivalent to the original ones. In general, the naivy addition of square terms in the equations of motion of
some given theory will lead to a physically different model.

In what follows, motivated by (\ref{la1k}) and (\ref{lnfp}), we use the addition of quadratic terms in the Euler
tensor as a technique to generalize the usual symmetric ($e_{\mu\nu}=e_{\nu\mu}=h_{\mu\nu}$) massive Fierz-Pauli
theory. We start with the general Ansatz :

\be {\cal L}_G [h_{\mu\nu}] = {\cal L}_{FP}[h_{\mu\nu}] +
\frac{a}2 \left( \p_{\mu}K^{\mu\nu}\right)^2 + b \,  \p_{\mu} K \,
\p_{\alpha}K^{\alpha\mu} + \frac{c}2 \left( \p_{\mu}\p_{\nu}
K^{\mu\nu}\right)^2 + \frac{d}2 K^2_{\mu\nu} + \frac{f}2 K^2 \no
\label{lg} \ee

\no where $K^{\mu\nu}$ is given in (\ref{K}) and  ${\cal L}_{FP}$
is defined in the two lines of (\ref{nsfp}). The coefficients
$a=a(\Box), b=b(\Box ), c=c(\Box), d=d(\Box) $ and $f=f(\Box )$
are in principle arbitrary analytic functions (Taylor series) of
$\Box=\p^{\mu}\p_{\mu}$. The equations of motion of the general
Ansatz can be cast in the form

\be K^{\mu\nu}_G = \frac{\delta \, S_G}{\delta \, h_{\mu\nu}} = K^{\mu\nu} + \hat{a}^{\mu\nu}_{\,\,\,\,
\alpha\beta} \, K^{\alpha\beta} = 0 \quad .  \label{eomsg} \ee

\no where $\hat{a}$ is a differential operator given in terms of the arbitrary coefficients of the Ansatz
(\ref{lg}). Notice that the FP equations of motion $K^{\mu\nu}=0$ are embedded in the set of solutions of
(\ref{eomsg}). By applying $\hat{a}$ on (\ref{eomsg}) we obtain (suppressing indices)

\be \hat{a}\, K + \hat{a}^2 \, K = 0 \quad . \label{a2} \ee

\no Our task is to fix the free coefficients in (\ref{lg}) such that $\hat{a}^2 \, K = 0 $. Consecutively, we
have on shell $\hat{a}\, K = 0$, back in (\ref{eomsg}) we deduce $K^{\mu\nu}=0$. So we prove the equivalence of
(\ref{eomsg}) and the original equations of motion of the massive Fierz-Pauli theory. Splitting according to the
tensor structure, we can write down explicitly

\bea \left(\hat{a}^2 \, K\right)^{\mu\nu} &=& d^2\left(\Box -
m^2\right)^2 K^{\mu\nu} + A_5 \left\lbrack 2\, d (\Box - m^2) +
A_5 \Box \right\rbrack \left(\p^{\mu}\p_{\alpha}K^{\alpha\nu} +
\p^{\nu} \p_{\alpha}K^{\alpha\mu} \right) \nn \\
&+& \eta^{\mu\nu} \left\lbrack j_1(\Box) \, R_L + j_2(\Box) \, h
 + \cdots \right\rbrack + \p^{\mu}\p^{\nu} \left\lbrack g_1(\Box) \, R_L +
g_2(\Box) \, h  + \cdots \right\rbrack \quad . \label{a2k} \eea

\no where $A_5 = m^2 a(\Box)/2 - d(\Box)$ while $R_L = \p^{\mu}\p^{\nu} h_{\mu\nu} - \Box \, h $ may be
interpreted as a linearized scalar curvature about a flat background ($g_{\mu\nu} = \eta_{\mu\nu} + h_{\mu\nu}
$). The dots stand for terms which vanish\footnote{The coefficients $a,b,c,d,f$ are always supposed to be
functions of $\Box$ unless otherwise stated.} at $A_5=0=d$. We keep the first line of (\ref{a2k}) for later
reference.  The functions $j_i(\Box) , g_i(\Box), i=1,2$ are given by

\bea j_1 (\Box) &=& - m^2 A_1 (A_1 \Box + A_2 \Box^2) - m^2 A_3
(D\, A_1 + A_2 \Box ) + m^2 \frac{(D-2)}{(D-1)} j_2(\Box) \label{f1} \\
j_2 (\Box) &=& m^2 (D-1) \left\lbrack A_1(A_3 \Box + A_4 \Box^2 )
+ A_3( D\, A_3 + A_4 \Box ) \right\rbrack \label{f2} \\
g_1 (\Box) &=& - m^2 A_2 (A_1 \Box + A_2 \Box^2) - m^2 A_4 (D\,
A_1 + A_2 \Box ) + m^2 \frac{(D-2)}{(D-1)} g_2(\Box) \label{g1} \\
g_2 (\Box) &=& m^2 (D-1) \left\lbrack A_2(A_3 \Box + A_4 \Box^2 )
+ A_4( D\, A_3 + A_4 \Box ) \right\rbrack \label{g2} \eea

\no where

\bea A_1 &=& \left\lbrack b (D-2) + c\, m^2 \right\rbrack \Box -
m^2 b \, (D-1) \quad ; \quad A_2 = - b (D-2) - c\, m^2 \quad ,
\label{a1a2} \\
A_3 &=& - \left\lbrack b \, m^2 + f (D-2) \right\rbrack \Box +
f(D-1)m^2  \quad ; \quad A_4 = b\, m^2 + f(D-2) \quad .
\label{a3a4} \eea

Regarding the tensor structure, the only terms proportional to $h^{\mu\nu}$ and
$\p^{\mu}\p_{\alpha}h^{\alpha\nu} + \p^{\nu}\p_{\alpha}h^{\alpha\mu} $ in (\ref{a2k}) come respectively from
$K^{\mu\nu}$ and $\p^{\mu}\p_{\alpha}K^{\alpha\nu} + \p^{\nu} \p_{\alpha}K^{\alpha\mu}$. Thus, in order that $
\left(\hat{a}^2 \, K\right)^{\mu\nu} =0$ we must have $d=0=A_5$ which is equivalent to $d=0=a$. The remaining
terms in $ \left(\hat{a}^2 \, K\right)^{\mu\nu}$ will vanish {\bf identically} if we require
$j_1=0=j_2=g_1=g_2$. The reader can check that there are only two solutions to those equations according to
$f=0$ or $f\ne 0$. We have respectively,

\be f=0=b \quad , \quad ({\rm Solution} \quad {\rm I})
\label{sol1} \ee

\be c = \frac{b^2}f \quad ; \quad b = \frac{f\left\lbrack m^2 \, D
- (D-2) \Box \right\rbrack }{2\, m^2 \, \Box} \quad , \quad ({\rm
Solution } \quad {\rm II} ) \label{sol2} \ee

In solution I, the coefficient $c(\Box)$ remains an arbitrary analytic function. In solution II the
arbitrariness lies in the coefficient $f(\Box)$ which can be any Taylor series which must start however, at the
second power $\Box^2$ by locality reasons as we will see later.

In the case of solution I the Ansatz (\ref{lg}) becomes :

\be {\cal L}^{I} = {\cal L}_{FP}[h_{\mu\nu}] + \frac 12 \left(\p^{\mu}\p^{\nu}h_{\mu\nu} - \Box \, h
\right)c(\Box)\left(\p^{\mu}\p^{\nu}h_{\mu\nu} - \Box \, h \right) \quad . \label{lfpc} \ee

\no The function $c(\Box)$ must have inverse mass squared dimension. The equations of motion $\delta S^{I} = 0 $
are given by

\be K^{\mu\nu} + c(\Box) \Box \theta^{\mu\nu}\Box
\theta^{\alpha\beta}h_{\alpha\beta} = 0  \quad . \label{eomI} \ee

\no where $\Box \theta^{\mu\nu} = \eta^{\mu\nu}\Box - \p^{\mu}\p^{\nu} $. Instead of using the operator
$\hat{a}$, defined in (\ref{eomsg}),  it is simpler to apply $\p_{\mu}\p_{\nu}$ on (\ref{eomI}). Only the mass
terms of (\ref{K}) will contribute and we get $\Box \theta^{\mu\nu}h_{\mu\nu} = 0$, back in (\ref{eomI}) we
recover the usual FP equations of motion $K^{\mu\nu}=0$. So, although the higher order term in (\ref{lfpc}) is
not a total derivative and contains more than two time derivatives, the new theory ${\cal L}^I$ is on shell
equivalent to the usual second order FP theory.

%
%

Remarkably, the theory $ {\cal L}^I$ has nonlinear completions. We can choose for instance


\be S^I[g_{\mu\nu},g_{\mu\nu}^{(0)}] = \frac 1{2\, \kappa^2} \int \, d^D x \,  \sqrt{-g}\left\lbrack  R + R\,
c(\Box) \, R  - \frac{m^2}4  (h_{\mu\nu}h^{\mu\nu} - h^2) \right\rbrack \, . \label{snl1}\ee

\no where $h_{\mu\nu}=g_{\mu\nu} - g_{\mu\nu}^{(0)}$ while
$g_{\mu\nu}^{(0)} $ is some fixed metric as opposed to the dynamic
one $g_{\mu\nu}$. In the last term of (\ref{snl1}) the indices of
$h_{\mu\nu}$ are raised with $g_{\mu\nu}$.

Expanding $S^I$ about flat space $g_{\mu\nu}^{(0)}= \eta_{\mu\nu}$
with  $g_{\mu\nu} = \eta_{\mu\nu} + h_{\mu\nu}$ we recover
(\ref{lfpc}) at quadratic order. The constant $\kappa$ has mass
dimension $1-D/2$.

 The
fixed metric breaks the general coordinate invariance of the
scalar curvature terms but it could be restored by introducing
\stuck fields in different ways, as explained in \cite{hinter} in
the $c=0$ case.

According to \cite{cs} the addition of fine tuned, as in
\cite{rgt}, higher powers of $h_{\mu\nu}$ (without derivatives)
may render the model (\ref{snl1}), with constant $c(\Box)=c_0$,
ghost free at nonlinear level. In \cite{cs} the cosmology of the
$R + c_0 R^2$ model, and of an $f(R)$ generalization thereof, has
been investigated leading to both early time inflation and late
time accelerated expansion.

The solution II corresponds to

\be {\cal L}^{II} = {\cal L}_{FP}[h_{\mu\nu}] + \frac{\Phi \,
f(\Box) \, \Phi }{8\, \Box^2} \quad . \label{lfpf} \ee

\no where

\be \Phi = \left\lbrack (D-2)\Box + D\, m^2 \right\rbrack R_L + 2\, m^2 (D-1)\Box \, h \quad . \label{phi} \ee

\no The analytic function  $f(\Box)$ is arbitrary except for the
fact that it must start at the second power $\Box^2$ as required
by locality.

The contribution of the last term of (\ref{lfpf}) to the equations of motion $\delta S^{II}/\delta h_{\mu\nu} =0
$ is proportional to the scalar $\Phi$. If we show that $\Phi =0$ on shell, then we prove the equivalence
between $\delta S^{II}/\delta h_{\mu\nu} =0 $ and $K^{\mu\nu}=0$. In order to show that $\Phi =0$ we can take a
general combination $s(\Box) \p_{\mu}\p_{\nu} + t(\Box) \eta_{\mu\nu}$ and apply on   $\delta S^{II}/\delta
h_{\mu\nu}=0$. It turns out that if we choose $s(\Box) = D\, m^2 - (D-2)\Box $ and $t(\Box) = - 2 \, m^2 \, \Box
$ we derive $\Phi =0 $. Consequently, once again, though the last term in (\ref{lfpf}) is not a total
derivative, the equations of motion of (\ref{lfpf}) are equivalent  to the usual FP equations $K^{\mu\nu}=0$.

The theory (\ref{lfpf}) also has nonlinear completions, for instance,

\be S^{II}[g_{\mu\nu},g_{\mu\nu}^{(0)}] = \frac 1{2\, \kappa^2}
\int \, d^D x \,\sqrt{-g}  \left\lbrack  R + \frac{\Phi \,
\tilde{f}(\Box) \, \Phi }{2}  - \frac{m^2}4 (h_{\mu\nu}h^{\mu\nu}
- h^2) \right\rbrack \, . \label{snl2}\ee

\no where $\tilde{f}(\Box )= f(\Box)/(8 \, \Box^2)$ is an arbitrary Taylor series in $\Box = \nabla_{\mu}
\nabla^{\mu}$ and $\Phi$ is defined as in (\ref{phi}) with the replacement, for instance, of $R_L (h)$ by the
full scalar curvature $R(g)$ and $\Box = \nabla_{\mu}\nabla^{\mu}$ . Moreover, $h_{\alpha\beta} =
g_{\alpha\beta} - g_{\alpha\beta}^{(0)} $ is a rank-2 tensor under general coordinate transformations. We assume
that $g_{\alpha\beta}^{(0)} $ has been ``covariantized'' into a rank-2 tensor under general coordinate
transformations via introduction of \stuck fields, as explained, e.g., in \cite{hinter}. The operator $\Box =
\nabla_{\mu}\nabla^{\mu}$  is a contraction of two covariant derivatives with respect to the metric
$g_{\mu\nu}$.  Expanding $S^{II}$ about flat space we recover (\ref{lfpf}).

We have obtained solutions I and II from the condition $\hat{a}^2 \, K = 0$. The first line of (\ref{a2k})
vanished by imposing $a=0=d$ while the the second line  vanished {\bf identically} by requiring
$j_1=0=j_2=g_1=g_2$. However, we could have instead a weaker condition where $\hat{a}^2 \, K = 0$ holds only on
shell, i.e., as a consequence of $K^{\mu\nu}_G = 0 $. One can convince oneself that the first line of
(\ref{a2k}) can not vanish on shell, so we still need to fix $a=0=d$. Regarding the second line, if we had
$\p^{\mu}\p^{\nu}h_{\mu\nu} = 0 = h$ as a consequence of the full equations of motion $K^{\mu\nu}_G = 0 $, there
would be no need of $j_1=0=j_2=g_1=g_2$. Indeed, assuming $a=0=d$, let us examine the scalar equations of
motion:

\bea \p_{\mu}\p_{\nu}K^{\mu\nu}_G &=& \left\lbrack 1 + d(\Box - m^2)+ A_1\Box + A_2 \Box^2 \right\rbrack \dk +
(A_3 + A_4 \Box ) \Box \, K  \nn \\
&=& P \, \llh + Q \, \Box \, h = 0 \quad . \label{pq} \eea

\bea \Box  \,  \eta_{\mu\nu} K^{\mu\nu}_G &=& \left\lbrack 1 + d(\Box - m^2)+ D \, A_3 + A_4 \Box \right\rbrack
\Box \, K
+ (D A_1 + A_2 \Box ) \, \Box \, \dk \nn \\
& = & S \, \llh + T \, \Box \, h  = 0  \quad . \label{rs} \eea

\no The equations  (\ref{pq}) and (\ref{rs}) define $P,Q,S,T$
unambiguously  as functions of $\Box$.  In order to have only
trivial solutions $\llh = 0 = \Box h$ the determinant $P T - Q S $
must be a nonvanishing constant. Explicitly,

\be m^4 (D-1)^2(b^2- c\, f)\Box^2 - (D-1)(2\, b \, m^2 + f (D-2))\Box + D(D-1)m^2 f + 1 = C \label{det} \ee

\no where $C = (P T - Q S)/[m^4(D-1)]$ is required to be a non
vanishing real constant thus leading to  $\llh = 0 = \Box h$. Back
in the scalar equation $\eta_{\mu\nu} K^{\mu\nu}_G = 0$ we deduce
the null trace $h=0$, thus guaranteeing $\hat{a}^2 \, K = 0$ on
shell.

There are several solutions to (\ref{det}). In particular, solutions I and II satisfy (\ref{det}) with $C=1$.
Another interesting solution corresponds to constant coefficients:

\be   b = - \frac{(D-2)\, f}{2 \, m^2} \quad ; \quad c =
\frac{(D-2)^2\, f}{4 \, m^4} \quad \quad {\rm ( Solution \quad III
)} \label{s3} \ee

\no It corresponds to $ C = 1 + D(D-1)\, f \, m^2 $ where $f$ is
an arbitrary real constant such that $C \ne 0$. The corresponding
Lagrangian differs from the Fierz-Pauli theory by a square term:

\be {\cal L}^{III} =  {\cal L}_{FP} + \frac f8 \left\lbrack
(D-2)R_L + 2 (D-1)\, m^2 \, h \right\rbrack^2 \quad . \label{l3}
\ee

\no The fact that the equations of motion $\delta \, S^{III} =0 $
lead to $K^{\mu\nu}=0$ directly follows from  the on shell
identity:

 \be \left\lbrack (D-2)\p_{\mu}\p_{\nu} + 2
\, m^2 \, \eta_{\mu\nu} \right\rbrack \frac{\delta S^{III}}{\delta
h_{\mu\nu}} = m^2 \left\lbrack 1 + f\, m^2 D(D-1) \right\rbrack
\left\lbrack (D-2)R_L + 2 (D-1)\, m^2 \, h \right\rbrack =0
\label{s3eom} \ee

\no Regarding a nonlinear completion for the third solution, we
may have for instance

\be S_{III} = \frac 1{2\, \kappa^2} \int \, d^D x \, \sqrt{-g}\,
\left\lbrace R  - \frac{m^2}4 (h_{\mu\nu}h^{\mu\nu} - h^2) + \frac
f{16} \, \left\lbrack (D-2)R + 2 (D-1)\, m^2 \, h \right\rbrack^2
\right\rbrace \, . \label{snl3}\ee

\no where $f$ is a real constant with inverse mass squared dimension such that $1 + D(D-1)\, f \, m^2 \ne 0 $.
Once again $h_{\mu\nu} = g_{\mu\nu} - g_{\mu\nu}^{(0)} $  and (\ref{l3}) is the quadratic truncation of
(\ref{snl3}) about flat space.

Remarkably, comparing with the usual massive theory (\ref{snl1}), in $S_{III}$ we have both an ultraviolet and
an infrared modification. The non derivative mass term now departures from the usual FP structure but remains
ghost free at quadratic level. In the usual FP case ($f=0$) one can get rid of the Boulware-Deser ghost at
nonlinear level by the addition of fine tuned non derivative terms \cite{rgt}. Since the IR and UV modifications
in (\ref{snl3}) are interconnected, we believe that possible ghost free modifications of \cite{rgt} must include
both non derivative and higher order terms.

In order to find and classify solutions to the determinant condition (\ref{det}) we should consider the
coefficients $b, f, c$ as Taylor series in $\Box$ and require that each power of $\Box$ in (\ref{det}) vanishes.
Although we are still struggling with the search of solutions to (\ref{det}), we have been able so far to find a
fourth class  of solutions (Solution IV):

\bea f &=& f_0 + f_1 \, \Box \quad ; \quad b = b_0 + b_1 \, \Box
\quad ; \quad c = c_0 + c_1 \, \Box  \nn\\
f_0 &=& 0 \, ; b_0 = D \frac{f_1}2 \, ; \, b_1 = \frac{f_1}{8\,
m^2}  \left\lbrack D^2(D-1) f_1 m^4 - 4(D-2)\right\rbrack \, ; \,
c_0 = D\, b_1 \, ; \, c_1 = \frac{b_1^2}{f_1}   \label{s4} \eea

\no The corresponding flat space Lagrangian is given by

\be {\cal L}^{IV} = {\cal L}_{FP} + \frac{f_1}{128} \left\lbrack N\, R_L + 8(D-1)m^2\, h \right\rbrack
\left\lbrace \Box \left\lbrack N\, R_L + 8(D-1)m^2\, h \right\rbrack + 8\, D \, m^2 \, R_L \right\rbrace \, .
\label{l4} \ee

\no where $N= D^2(D-1)m^4 f_1 + 4 (D-2) $ is an arbitrary real
number which follows from the arbitrariness of $f_1$. Notice that
$f_1$ is not just an overall factor  since $N=N(f_1)$. An
interesting subcase, we call solution IV-a, corresponds to choose
$f_1$ such that $N=0$. We end up with a second order theory which
differs from the usual massive FP theory by a trivial field
redefinition (Weyl transformation):

\be {\cal L}^{IV-a}[ h_{\mu\nu}] = {\cal L}_{FP}[h_{\mu\nu}] + 2 \frac{(D-2)}D \p^{\mu}h \,
\p^{\alpha}h_{\alpha\mu} - 2 \frac{(D-2)}{D^2} \p^{\mu} h \, \p_{\mu} \, h = {\cal L}_{FP}[h_{\mu\nu} -
(2/D)\eta_{\mu\nu} h ]\quad . \label{l4a} \ee

\no The Weyl transformation $ h_{\mu\nu} \to h_{\mu\nu} -
(2/D)\eta_{\mu\nu} h $ is the only one which preserves the form of
the FP mass term. Even though ${\cal L}^{IV-a}$ differs trivially
from the FP theory at quadratic level, a nonlinear completion of
IV-a may be quite different from (\ref{snl1}) with $c=0$, for
instance, we may have

\be S^{IV-a} [g_{\mu\nu},g_{\mu\nu}^{(0)}] = \frac 1{2\, \kappa^2}
\int \, d^D x \,  \sqrt{-g}  \left\lbrace R - \frac{m^2}4
 (h_{\mu\nu}h^{\mu\nu} - h^2)
 + \frac{(D-2)}{D^2} \, h \left\lbrack \Box \, h -
D\, \nabla^{\mu} \nabla^{\nu} h_{\mu\nu} \right\rbrack \right\rbrace \label{s4a}\ee

\no  Expanding $S^{IV-a} $ around flat space
($g_{\mu\nu}=\eta_{\mu\nu} + \, h_{\mu\nu}$) we recover
(\ref{l4a}) at quadratic level. According to \cite{rmt13}, where a
complete set of cubic and quartic (in powers of $h_{\mu\nu}$)
vertices with two derivatives has been analyzed, there is
apparently no hope of a ghost free (at nonlinear level) second
order model.

In the general case $N \ne 0$ we may have the nonlinear fourth order completion:

\be S_{IV} = \frac 1{2\, \kappa^2} \int \, d^D x \,
\sqrt{-g}\left\lbrace
 R  - \frac{m^2}4  (h_{\mu\nu}h^{\mu\nu}
- h^2) + \frac {f_1}{264} \,  \Psi \, \left( \Box \, \Psi + 8\, D
m^2\, R \right) \right\rbrace . \label{s4}\ee

\no where  we have introduced the curved space scalar $ \Psi = N\,
R + 8(D-1)m^2\, h $.

There are other Taylor series $f(\Box), b(\Box), c(\Box)$ which solve the determinant condition (\ref{det}). We
leave for a future work the detailed analysis of those solutions altogether with a study of possible ghost-free
models at nonlinear level.

%
%
%
%
%

\section{The propagator}

In order to confirm that the higher order (in derivatives)
generalizations of the Fierz-Pauli theory suggested in the last
section are indeed ghost  and tachyon free, we examine here the
analytic structure of the propagator of those theories. In
general, a Lagrangian of the form (\ref{lg}) leads to the
propagator (suppressing indices)

\bea G^{-1} &=& \frac{2\, P_{SS}^{(2)} }{\left( \Box - m^2 \right)
\left\lbrack d (\Box - m^2)+1\right\rbrack } - \frac{2\,
P_{SS}^{(1)}}{m^2 \left( 2 + a\, m^2 \Box - 2 \, d\, m^2 \right)}
\nn\\ &+& \frac{ \left\lbrack A_{ww}\, P_{WW}^{(0)} + A_{ws}
\left( P_{SW}^{(0)} + P_{WS}^{(0)}\right) +
A_{ss}P_{SS}^{(0)}\right\rbrack } {K^{(0)}} \label{propa1} \eea

\no where the determinant $K^{(0)}=A_{ww}A_{ss}-A_{ws}^2 $ is
given in terms of the coefficients $A_{ij}=A_{ij}(\Box)$  which
are complicated functions of the coefficients $a,b,c,d,f$  given
in the appendix. The spin-s operators $P_{IJ}^{(s)}$ are also
displayed in the appendix.

Now we can understand why we need to have $d=0=a$ from a different
viewpoint. Namely, if $d \ne 0$ we would have a double pole in
spin-2 sector which would necessarily lead to a spin-2 ghost. We
need to chose $a=0$ to avoid a massive vector particle. With
$d=0=a$ the spin-2 and spin-1 sectors of the propagator
(\ref{propa1}) coincide with the corresponding sectors of the
propagator  of the usual second order FP theory. In the spin-0
sector it turns out that $K^{(0)} = -(P T - Q S)/4 = - m^4(D-1)C/4
$ where $C$ is defined by the left hand side of (\ref{det}). Thus,
the requirement of last section that $C$ be a nonvanishing real
constant is equivalent to the absence of poles in the spin-0
sector of the propagator. Therefore there is a simple
interpretation of the Euler tensor approach in terms of the
analytic structure of the propagator. Namely, it amounts to
preserve the analytic structure of the usual FP propagator.

\section{Euler tensor approach versus gauge embedment}

Since there is another approach in the literature \cite{anacleto,
clovis, sd4} where dual (higher order) models are generated by the
addition of quadratic terms in the Euler tensor, it is worth
pointing out the essential differences between those techniques.

 As shown\footnote{The first proof of equivalence between the Maxwell-Chern-Simons theory of
 \cite{djt} and the first order Self-dual model of \cite{tpn} has appeared in \cite{dj},
  but for our purposes the work \cite{anacleto} is more convenient.}
  in \cite{anacleto}, one can obtain in $D=2+1$  the second order,
 Maxwell-Chern-Simons (MCS) theory of \cite{djt}
via a Noether gauge embedment (NGE) of the first order self-dual
(SD) model of \cite{tpn}. Namely,

\be {\cal L}_{MCS} = \frac 12 f_{\mu}K^{\mu} + \frac 1{2\, m^2}
K_{\mu} K^{\mu} = {\cal L}_{SD} +   \frac 1{2\, m^2} K_{\mu}
K^{\mu} \quad . \label{mcs} \ee

\no Defining the transverse operator $E_{\mu\nu} \equiv
\epsilon_{\mu\nu\alpha}\p^{\alpha}$, the Euler vector $K_{\mu}$ of
the SD model is given by  $K_{\mu}= \delta \, S_{SD}/\delta
f^{\mu} = m \, E_{\mu\nu}f^{\nu} - m^2 f_{\mu} $. An arbitrary
variation of the MCS action

\be \delta S_{MCS} = \int \, d^3 x \, \left\lbrack K_{\mu} \delta
\, f^{\mu} + \frac{K_{\mu}}{m^2} \left( m \, E^{\mu}_{\,\,\,\,
\nu}\delta f^{\nu} - m^2 \, \delta f^{\mu} \right) \right\rbrack =
\frac 1m \int \, d^3 x \, K^{\mu} E_{\mu\nu} \delta f^{\nu}  \quad
. \label{dmcs} \ee

 \no reveals the $U(1)$ symmetry $\delta
f_{\mu} = \p_{\mu} \phi $ of the MCS  theory, contrary to the SD
model which has no local symmetry.

Moreover, (\ref{dmcs}) reveals also that the SD equations of
motion $K_{\mu}=0$ are embedded in the MCS ones $E_{\mu\nu}K^{\nu}
= 0$. Similarly, the linearized new massive gravity (NMG) of
\cite{bht}, which is of fourth order in derivatives, can be
obtained from the usual second order massive FP theory by adding
quadratic terms in the FP Euler tensor, see \cite{sd4}. The
linearized NMG theory is invariant under $\delta h_{\mu\nu} =
\p_{\mu} \zeta_{\nu} + \p_{\nu} \zeta_{\mu} $ contrary to the FP
theory which has no local symmetry. In both cases, SD/MCS and
FP/NMG, we have a Noether gauge embedment.

Contrary to the Euler tensor approach of section 2, in the NGE it
is not possible to prove that the new (higher order) equations of
motion imply the old ones. In fact, this is not expected on
general grounds since the old equations, like $K_{\mu} = 0$, are
not gauge invariant. The on shell equivalence between the dual
theories is more subtle in the NGE. For instance, in the SD/MCS
case one defines a gauge invariant dual field $F_{\mu} =
E_{\mu\nu}f^{\nu}/m $ and shows that the MCS equations of motion
can be cast in the SD {\bf form} $K_{\mu} = 0$ with $f_{\mu} \to
F_{\mu}$. A similar dual map exists also in FP/NMG case, see
\cite{sd4}.

There is a crucial difference also in the analytic structure of the propagator if we compare the Euler tensor
 approach of last section to the NGE. In the former, the changes in the propagator do not affect its analytic structure
 at all. The highest spin sectors remain unchanged while the contribution of the
different spin-0 terms that we have introduced in the action have
 a vanishing net influence in $K^{(0)}$. On the other hand, in
the NGE there appear new poles in the denominators even in the
highest spin (physical) sector. However, when we saturate the
two-point amplitude with sources satisfying the constraints
imposed by the larger symmetries, the new poles have a vanishing
residue, see \cite{oda,unitary}. This already happens in the
SD/MCS duality. In the MCS model there is a massless pole in the
propagator, with vanishing residue,  which is not present in the
SD model. This is how the NGE solves the puzzle of having higher
derivative theories while preserving unitarity\footnote{We thank
Prof. Jos\'e A. Helayel-Neto for a comment on that point.}. It is
a rather different solution as compared to the Euler tensor
approach where the different higher order terms in the action  are
fine tuned in order to produce no contributions at all in the
denominators in the propagator, thus avoiding extra poles from the
start.

\section{Conclusion}

Here we have made embeddings of the Euler tensor as a technique to produce dual theories to the massive
Fierz-Pauli model. The models obtained are in general of higher order in derivatives but remain unitary as
demonstrated via analysis of the analytic structure of the corresponding propagators. The Euler tensor approach
could be in principle generalized to other models. The coefficients appearing in the Ansatz (\ref{lg}) are in
general analytic functions of $\Box = \p^{\mu}\p_{\mu}$ which must satisfy condition (\ref{det}). There are
several solutions to (\ref{det}). We have written down explicitly four set of solutions. In all cases the
(usually) higher order equations of motion  turn out to be equivalent to the second order Fierz-Pauli equations,
thus leading to the Fierz-Pauli conditions and the Klein-Gordon equation. They describe a massive spin-2
particle if $D=4$.

All models admit nonlinear completions in terms of two metrics,
$g_{\mu\nu}$  and  $g_{\mu\nu}^{(0)}$, see
(\ref{snl1}),(\ref{snl2}), (\ref{snl3}) and (\ref{s4}). The
solutions I and II are all of higher order in derivatives. The
special subcase of solutions I, see (\ref{snl1}), where $c(\Box)$
is a constant has appeared before in \cite{cds} where, following
\cite{rgt},  nonlinear non derivative terms have been added to
render the model ghost free at nonlinear level. The authors of
\cite{cds} have found interesting self-accelerating solutions.

For the solutions III and IV the situation is different. In solutions III the usual Fierz-Pauli (non derivative)
mass term gets modified altogether with a higher order $R^2$ term and a second order  modification of the
Fierz-Pauli theory, see (\ref{snl2}). So we believe that the addition of nonlinear non derivative terms must
come together with nonlinear higher derivative terms in order to get rid of the Deser-Boulware ghost eventually.
In the set of solutions IV once again the second order Fierz-Pauli terms must change along with higher order
terms, thus requiring a more careful study of possible ghost free additions.

It is possible to lower the order (in derivatives) of (\ref{snl2}) and (\ref{snl3}) by introducing a scalar
field. Since the use of a scalar field seems to be an important ingredient in the proof of absence of ghosts, at
nonlinear level, in the higher order theories of  \cite{cs}, there is some hope of a generalization.

We are currently investigating the generalization of the Ansatz (\ref{lg}) to the nonsymmetric case, where we
start with the nonsymmetric FP theory (\ref{nsfp}) and add quadratic terms in the nonsymmetric Euler tensor
$\tk^{\mu\nu}$.

\section{Appendix}

Here we display the operators $P_{IJ}^{(s)}$ and the coefficients
$A_{ij}(\Box)$  mentioned in section 3.

 Using, as building blocks the spin-0 and spin-1 projection
operators acting on vector fields, respectively,

\be  \omega_{\mu\nu} = \frac{\p_{\mu}\p_{\nu}}{\Box} \quad , \quad
\theta_{\mu\nu} = \eta_{\mu\nu} -
\frac{\p_{\mu}\p_{\nu}}{\Box}\quad , \label{pvectors} \ee

\no  we define the spin-s operators $P_{IJ}^{(s)}$ acting on
symmetric rank-2 tensors in $D$ dimensions:

\be \left( P_{SS}^{(2)} \right)^{\lambda\mu}_{\s\s\alpha\beta} =
\frac 12 \left( \theta_{\s\alpha}^{\lambda}\theta^{\mu}_{\s\beta}
+ \theta_{\s\alpha}^{\mu}\theta^{\lambda}_{\s\beta} \right) -
\frac{\theta^{\lambda\mu} \theta_{\alpha\beta}}{D-1} \quad ,
\label{ps2} \ee

\be \left( P_{SS}^{(1)} \right)^{\lambda\mu}_{\s\s\alpha\beta} =
\frac 12 \left(
\theta_{\s\alpha}^{\lambda}\,\omega^{\mu}_{\s\beta} +
\theta_{\s\alpha}^{\mu}\,\omega^{\lambda}_{\s\beta} +
\theta_{\s\beta}^{\lambda}\,\omega^{\mu}_{\s\alpha} +
\theta_{\s\beta}^{\mu}\,\omega^{\lambda}_{\s\alpha}
 \right) \quad , \label{ps1} \ee

\be \left( P_{SS}^{(0)} \right)^{\lambda\mu}_{\s\s\alpha\beta} =
\frac 1{3} \, \theta^{\lambda\mu}\theta_{\alpha\beta} \quad ,
\quad \left( P_{WW}^{(0)} \right)^{\lambda\mu}_{\s\s\alpha\beta} =
\omega^{\lambda\mu}\omega_{\alpha\beta} \quad , \label{psspww} \ee

\be \left( P_{SW}^{(0)} \right)^{\lambda\mu}_{\s\s\alpha\beta} =
\frac 1{\sqrt{D-1}}\, \theta^{\lambda\mu}\omega_{\alpha\beta}
\quad , \quad  \left( P_{WS}^{(0)}
\right)^{\lambda\mu}_{\s\s\alpha\beta} = \frac 1{\sqrt{D-1}}\,
\omega^{\lambda\mu}\theta_{\alpha\beta} \quad , \label{pswpws} \ee

\no They satisfy the symmetric closure relation

\be \left\lbrack P_{SS}^{(2)} + P_{SS}^{(1)} +  P_{SS}^{(0)} +
P_{WW}^{(0)} \right\rbrack_{\mu\nu\alpha\beta} =
\frac{\eta_{\mu\alpha}\eta_{\nu\beta} +
\eta_{\mu\beta}\eta_{\nu\alpha}}2 \quad . \label{sym} \ee

The coefficients $A_{ij}(\Box)$ are given by

\be  A_{ss} = z + y - (c_1 + c_2)\Box + c_3 \Box^2 \quad , \quad A_{ww} = z + y (D-1) \quad , \quad A_{ws} =
\sqrt{D-1}(c_2\Box/2 -y ) \quad . \ee

\no where

\bea c_1 &=& 1 + a \frac{m^4}2 + d(\Box - 2 \, m^2) \quad ,
\label{c1} \\
c_2 &=& -1 + d(D-2)(\Box - m^2) + \left\lbrack d + 2\, m^2 b (D-2)
+ c \, m^4 + f (D-2)^2 \right\rbrack \Box \nn\\
&-& m^2 (D-1)[b \, m^2 + f (D-2)] - a\, m^4 \quad . \label{c2} \\
 c_3 &=&  \frac {d \, D}2 + m^2 b (D-2) + \frac{c\, m^4}2 + \frac f2 (D-2)^2
\quad . \label{c3} \\
 z &=& \frac{(\Box - m^2)}2 \left\lbrack d (\Box - m^2) +
1\right\rbrack \quad . \label{z} \\
 y &=& - \frac{(\Box - m^2)}2 + d \,
\Box \left\lbrack \frac{(D-1)}2 \Box + m^2(3-D) \right\rbrack \nn
\\ &+& m^2 \left\lbrack b (D-2) + c \, \frac{m^2}2 + \frac{f(D-2)^2}{2
\, m^2} \right\rbrack \Box^2 + \frac f2 m^4 (D-1)^2 \nn \\ &-& m^2
(D-1)\left\lbrack b\, m^2 + f (D-2) \right\rbrack \Box \quad .
\label{w} \eea

\section{Acknowledgements}

The work of D.D. is supported by CNPq (307278/2013-1) and FAPESP (2013/00653-4) while A.L.R.S. is supported by
Capes. D.D. thanks Kurt Hinterbichler for a discussion on the model (\ref{s4a}).

\end{document}